# The Latency Validation of the Optical Link for the ATLAS Liquid Argon Calorimeter Phase-I Trigger Upgrade

Binwei Deng, [a,b] Le Xiao, [b,c] Xiandong Zhao, [b] Emily Baker, [d] Datao Gong, [b] Di Guo, [b,c] Huiqin He, [e] Suen Hou, [f] Chonghan Liu, [b] Tiankuan Liu, [b*] Quan Sun, [b] James Thomas, [b] Jian Wang, [b,g] Annie C. Xiang, [b] Dongxu Yang, [b,g] Jingbo Ye, [b] Wei Zhou [c]

[a] *Department of Electronic Information Engineering, Hubei Polytechnic University, Huangshi, Hubei 435003, P. R. China*

[b] *Department of Physics, Southern Methodist University, Dallas, TX 75275, USA*

[c] *Department of Physics, Central China Normal University, Wuhan, Hubei 430079, P.R. China*

[d] *Keller High School, Keller, TX 76248, USA*

[e] *Shenzhen Polytechnic, Shenzhen 518055, China*

[f] *Institute of Physics, Academia Sinica, Nangang 11529, Taipei, Taiwan*

[g] *State Key Laboratory of Particle Detection and Electronics, University of Science and Technology of China, Hefei Anhui 230026, China*

   *E-mail*: tliu@mail.smu.edu

ABSTRACT: Two optical data link data transmission Application Specific Integrated Circuits (ASICs), the baseline and its backup, have been designed for the ATLAS Liquid Argon (LAr) Calorimeter Phase-I trigger upgrade. The latency of each ASIC and that of its corresponding receiver implemented in a back-end Field-Programmable Gate Array (FPGA) are critical specifications. In this paper, we present the latency measurements and simulation of two ASICs. The measurement results indicate that both ASICs achieve their design goals and meet the latency specifications. The consistency between the simulation and measurements validates the ASIC latency characterization.

**Contents**

## 1. Introduction

The ATLAS detector at the Large Hadron Collider (LHC) has been successfully operating for almost a decade [1-2]. When the luminosity of LHC increases, the trigger system of the ATLAS detector must also be upgraded in order to reject the background more efficiently. The trigger readout electronics of the ATLAS Liquid Argon (LAr) Calorimeter is one part of the ATLAS Phase-I upgrade [3].

In the ATLAS LAr Calorimeter Phase-I trigger upgrade, the detector data are amplified, shaped, and digitized before they are transmitted to the back-end counting room through about 5000 optical fibers. The digitized data must be encoded and serialized before transmission. The functions of encoding and serializing are implemented in a device called the transmitter. Due to the harsh radiation environment [4-7], the transmitter is implemented in an Application Specific Integrated Circuit (ASIC).

For the ATLAS LAr Phase-I trigger upgrade, two ASICs, LOCx2 [8] and LOCx2-130 [9], are designed. LOCx2 is the baseline design and LOCx2-130 is a pin-compatible backup of LOCx2. LOCx2 is designed in a commercial 250-nm Silicon-on-Sapphire CMOS technology, whereas LOCx2-130 is designed in a commercial 130-nm bulk CMOS technology. LOCx2-130 uses a silicon-proven modified analog core, including a serializer and a Phase-Locked Loop (PLL), of GBTX [10] and TDS [11]. Both LOCx2 and LOCx2-130 are dual-channel transmitter ASICs. Each channel has a custom encoder and a serializer. The two channels share a PLL and a slow-control Inter-Integrated Circuit (I$^2$C) slave. In the back end, the deserializer and the decoder are implemented in a commercial Field-Programmable Gate Array (FPGA). Figure 1 is a simplified block diagram of LOCx2 and LOCx2-130 used in optical links.

The input data of LOCx2 and LOCx2-130 come from analog-to-digital converters (ADCs). LOCx2 and LOCx2-130 support three types of ADCs. All three types of ADCs sample analog signals at the LHC bunch crossing frequency of 40 MHz, but have a different resolution and number of analog channels. The first type of ADCs is an ASIC called the Nevis ADC [12]. Its resolution is 12 bits ($D_0$-$D_{11}$). However, its output data contain two calibration bits ($D_{12}$-$D_{13}$) and two dummy bits ($D_{14}$-$D_{15}$) in each conversion period, the LHC bunch crossing clock cycle of 25 ns. Each Nevis ADC has four analog channels. The other two types are Commercial-Off-The-Shelf (COTS) ADCs, ADS5272 and ADS5294, both produced by Texas Instruments [13]. Both COTS ADCs have eight analog channels. The resolutions are 12 bits and 14 bits for ADS5272 and ADS5294, respectively. The interface between ADCs and LOCx2/LOCx2-130 is similar. Each analog channel has a serialized digital data output (Channels 0-7) and all channels share a serial data clock (SCLK) and a frame clock (FCLK). The serial data clock can be used to latch the serialized data. The rising edge of the frame clock indicates the beginning of a frame, which is defined as the data transmitted in an LHC bunch-crossing clock cycle of 25 ns. Figure 2 shows the input data format of LOC2/LOCx2-130 coming from four Nevis ADCs.

The output serial data rates are 5.12 Gbps and 4.80 Gbps for LOCx2 and LOCx2-130, respectively. Figure 3 is the frame definition of the output serial data [9, 14]. The frame lengths are 128 bits and 120 bits for LOCx2 and LOCx2-130, respectively. Each frame consists of the payload, an optional Cyclic Redundancy Checking (CRC) field, and a frame trailer. The payload has 96 bits (for Nevis ADCs without calibration data and ADS5272) or 112 bits (for Nevis ADCs with calibration data and ADS5294). LOCx2 has an 8-bit CRC field, whereas LOCx2-130 has either a 16-bit CRC field or no CRC field. The frame trailer consists of a fixed pattern of 1010, which indicates the frame boundary, and a Pseudo-Random Binary Sequences field, which presents the frame index. The payload is scrambled [9, 14-15] before transmission to achieve DC balance, but neither the CRC field nor the frame trailer is scrambled.

The latency is a critical specification for LOCx2 and LOCx2-130 because of the constraint from the existing sub-detectors in the trigger system [2]. The latency of the ASIC is the time from the frame start (i.e., the rising edge of FCLK) of the input data to the frame start (i.e., the ending of the previous frame trailer) of the corresponding output data. The latency of the whole link (not including time passing through the optical fibers) is the time from the start of an input frame to the start of the recovered output frame. The latency budget of the ASIC and that of the whole link are 75 ns and 150 ns, respectively. Note the total latency of the optical transmitter/receiver is measured to be less than 1 ns, much less than those of other functional blocks, and is ignored in the following discussion. The latency of the ASICs has a variation of no larger than 3.125 ns with different ADC types. In this paper, we focus on Nevis ADCs in the nominal data mode.

In this paper, we present the latency measurements of LOCx2 and LOCx2-130, including the latencies of the ASICs and their corresponding receivers implemented in the FPGA, and the validation with FPGA simulation.

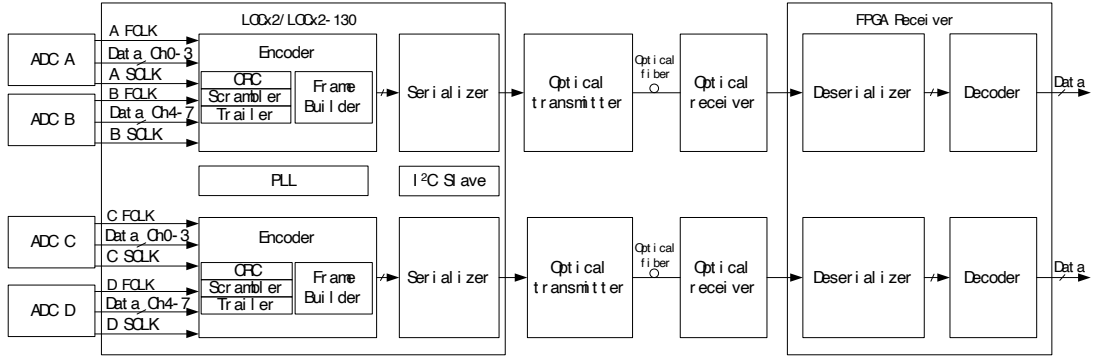

Figure 1: Block diagram of LOCx2/LOCx2-130 used in optical data links.

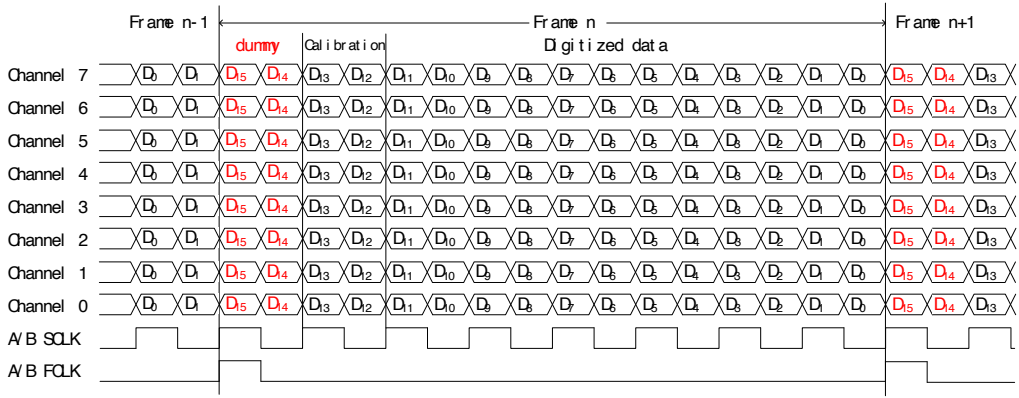

Figure 2: Input data format of LOCx2/LOCx2-130 from Nevis ADCs.

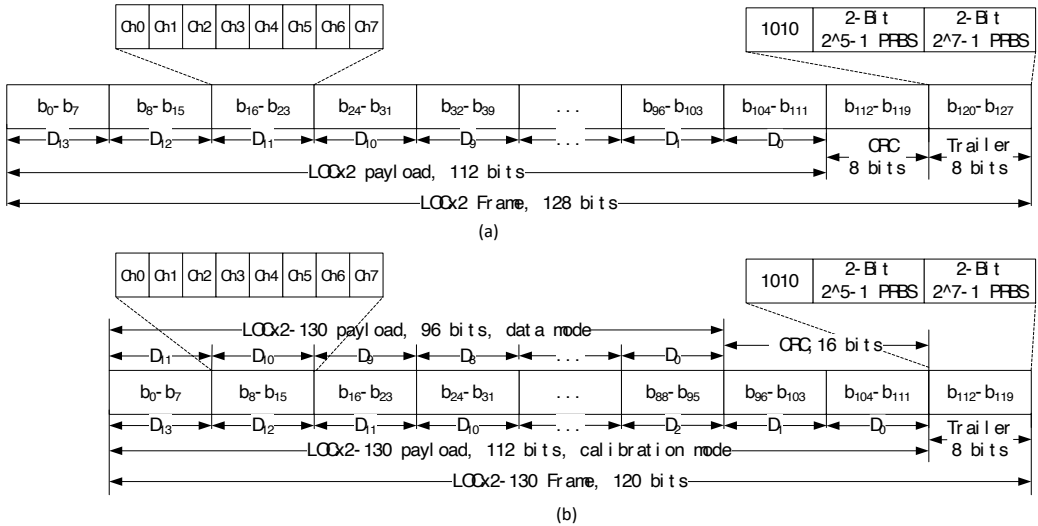

Figure 3: Frame definition of LOCx2 (a) and LOCx2-130 (b).

## 2. ASIC latency measurements

We obtained the ASIC latency by measuring the time from the input ADC data to the output. The measurement setup is shown in Figure 4. An FPGA evaluation board (Model No. KC705 produced by Xilinx) emulated four Nevis ADCs to generate their output signals. We aligned the output signals of all ADCs by using the I/O delay cells of the FPGA. There were 0.56-m twisted pairs of cables between the FPGA and the ASIC. We observed the ADC output signal FCLK and the high-speed serial data simultaneously by using a high-speed real-time oscilloscope (Model No. TDS72004 produced by Tektronix) with an active differential probe P7350 and a differential probe P7380SMA, respectively. The propagation delays of probes and cables were calibrated before the measurement. An SI5338 evaluation board generated a 320-MHz clock and a 40-MHz clock for the FPGA and the ASIC chip, respectively.

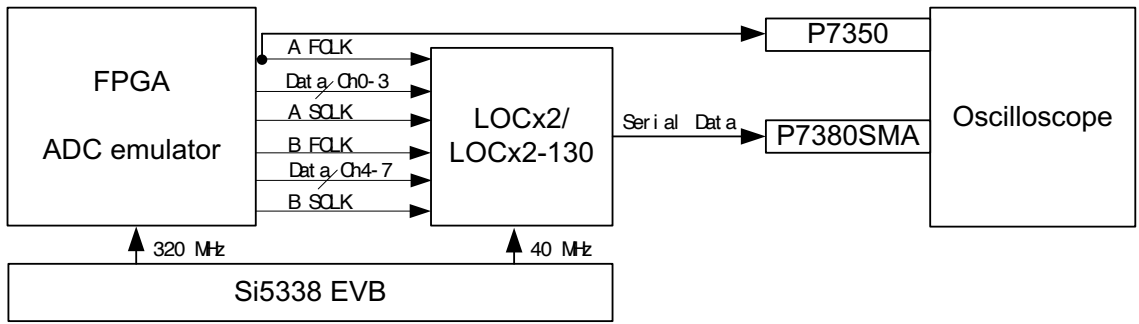

Figure 4: Block diagram of latency measurement setup.

The screenshot of the LOCx2-130 latency measurement is shown in Figure 5. The rising edge of an FCLK signal, measured at the output of the FPGA and marked as Cursor A in the figure, indicates the beginning of an input frame. When using such an FCLK signal as the trigger, due to the scrambling process, we cannot identify which bit in the high-speed serial data output is coming from which bit in the input ADC data except for the bits corresponding to the beginning of a frame trailer 1010, marked as Cursor B in the figure. The latency of the ASIC can be calculated from the time difference between the two cursors, after the trailer length and the propagation delays going through various cables and probes are taken into account. The time difference between the two cursors is 15.2 ns. The propagation delay of the 0.56-m twisted-pair cables between the FPGA and the ASIC is estimated to be 2.8 ns. The delay of the 1-m differential coaxial cables between the ASIC and the P7380SMA Probe is 4.8 ns, based on the datasheet of the cables. Relative to the P7380SMA probe, the P7350 probe has an additional 4.8-ns delay. Since there is a frame trailer in every 25 ns, the latency measurement has an ambiguity of $n \cdot 25$ ns (n = 0, 1, 2, 3…). We eliminate such an ambiguity by comparing to the ASIC simulation and the latency measurement. According to the ASIC simulation, the latency of LOCx2-130 is from 34.4 ns to 40.7 ns, which will be further explained in Section 4. The latency that is the closest to such a range is calculated to be 15.2 ns (the measured time difference between cursors) + 1.7 ns (the frame trailer length) + 4.8 ns (the extra delay of P7350 relative to P7380) - 2.8 ns (the delay of 0.56-m twisted-pair cables) - 4.8 ns (the delay of the 1-m coaxial cables) + 25 ns (frame length) = 39.1 ns. Such a latency measurement will be validated in the whole link latency measurement. The latency measurement of LOCx2 is similar to that of LOCx2-130.

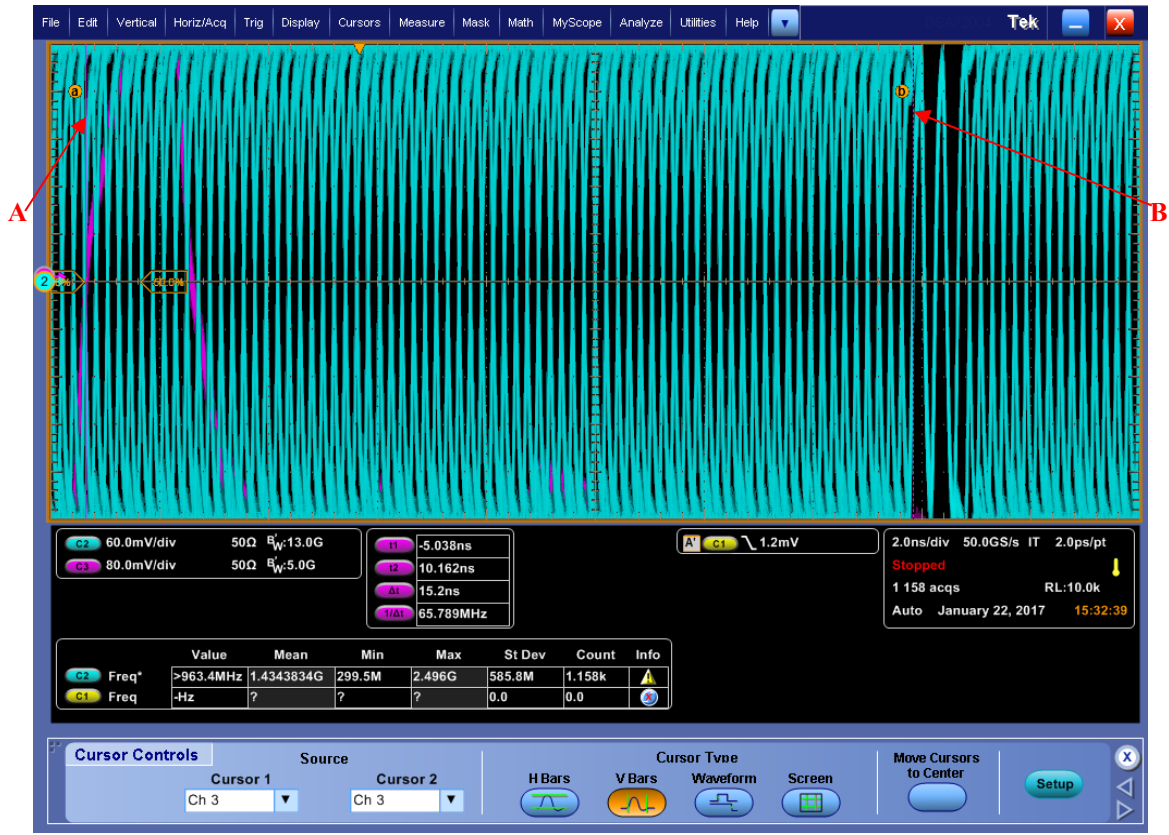

Figure 5: Screen shot of the LOCx2-130 latency measurement. The measured time between the two cursors is 15.2 ns.

**3. Latency of whole link**

We measured the whole link latency by comparing the time difference between the ADC output signals to the FPGA recovered signals. Figure 6 is the latency measurement setup of the whole link. An ADC emulator, which was implemented in an FPGA evaluation board, generated the ADC output signals. These ADC output signals were transmitted to the ASIC through 0.56-m twisted-pair cables. The high-speed serial output data of the ASIC were looped back to the FPGA through 1-m differential coaxial cables and recovered. In order to obtain the latency, each ADC emulator generated fixed pattern data in consecutive 255 frames and a special pattern in the following single frame. The data of Channels 7-0 in Figure 2 are 10101010 in 255 consecutive frames and 11001100 in the following single frame. The ADC emulator also generated a signal TxPulse aligned with the fixed pattern data 0xC. The receiver recovered the data and generated a signal RxPulse aligned with the recovered special pattern. The signals TxPulse and RxPulse were logically OR'ed in the FPGA and the output was sent to a real-time oscilloscope. The latency is the time difference between two rising edges of the OR'ed signal. The result for LOCx2 is shown in Figure 7. The two cursors indicate the positions of TxPulse and RxPulse. The time difference between the two cursors is 79.4 ns. The latency of the whole link is calculated to be 79.4 ns – 1.2 ns (the delay of the FPGA I/O delay cells) – 1.5 ns (the delay from the FPGA I/O delay cells to the FPGA output pins) – 2.8 ns (the delay of the 0.56-m twisted-pair cables from the FPGA to the ASIC) – 4.8 ns (the delay of the 1-m coaxial cables from the ASIC back to the FPGA) + 3.1 ns (the latency of CRC examiner) = 71.4 ns. The reason to add the time of the CRC examiner is that our RxPulse is generated simultaneously with the CRC examiner and the latency of the CRC examiner is not included in the measured latency. The measured latency falls in the simulated latency range from 68.2 to 74.3 ns, which will be described in the next section. For LOCx2-130, the measurement approach is similar and will not be described in detail.

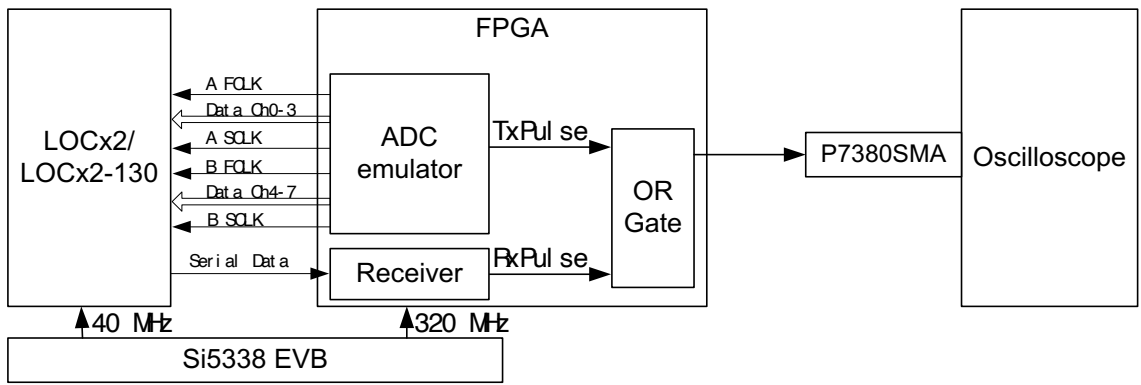

Figure 6: Setup for the latency measurement of the whole data link.

## 4. Latency validation

In order to resolve the n·25-ns uncertainty in the measured latency of the ASICs introduced in Section 2 and to obtain the latency of each functional block of the whole link, we validated the measured latencies in two approaches. The first approach is simulation. In addition to the simulation conducted for the ASIC design, we implemented both the transmitter and receiver sides in an FPGA using the ISim tool provided by Xilinx to simulate the whole link. The second approach uses the ChipScope Pro logic analyzer tool provided by Xilinx. For validation purpose, we used the second approach as much as possible. We used the Xilinx KC705 Evaluation Kit to implement the whole link.

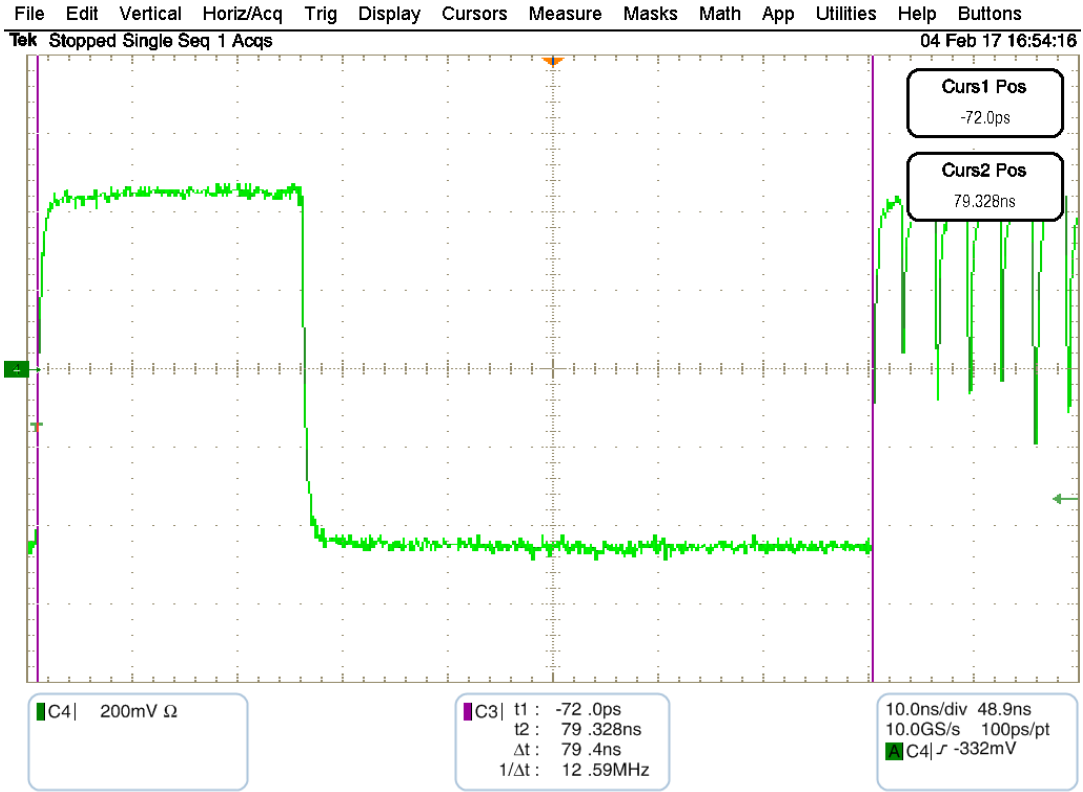

Figure 7: Whole link latency result for LOCx2. The measured time between the two cursors is 79.4 ns.

Based on simulation, the latency of each functional block of LOCx2 and LOCx2-130 as well as the latency of the whole link are listed in Table 1. The latency of the link implemented in the FPGA is also listed in the table for comparison. The consistency between the measured latency and simulation validates the ASIC latency characterization. In this section, we take LOCx2-130 as the example to describe how the latency validation was performed. The latency validation of LOCx2 is similar to that of LOCx2-130 and will not be described in detail.

Since the Kintex-7 FPGA did not support a 30:1 serializer, we implemented a 30-to-20 converter, which converted parallel 30-bit data each at the data rate of 160 Mbps to parallel 20-bit data each at the data rate of 240 Mbps. The recovered data on the receiver side were 20 bits each at 240 Mbps. In order to compare the data on the transmitter side and on the receiver side, we also generated the 20-bit data each at 240 Mbps on the transmitter side. We used a 20-to-30 converter to convert the 20-bit data for the encoder. The block diagram of the whole data link validation is shown in Figure 8(a). Figure 9 shows the validation results. The latency from the TxPulse, which is aligned with the special input pattern of the encoder, to the RxPulse, which is aligned with the recovered output special pattern, is 104.2 ns, or 25 clock cycles of 240 MHz. Considering the 0.3-m differential coaxial cables between the transmitter and the receiver, we calculate the latency of the whole link to be 104.2 ns – 1.4 ns (the delay of 0.3-m differential coaxial cables) = 102.8 ns. The measured latency falls in the simulated latency range from 100.8 to 104.8 ns, shown in Table 1. Note that the FIFO is not implemented in the FPGA. Since our RxPulse is generated simultaneously with the CRC examiner, the latency of the CRC examiner is not included in the whole link latency shown in the table.

In the block diagram of Figure 8(a), neither the 20-to-30 converter nor the 30-to-20 converter are a part of the real link and their latencies must be subtracted. For this purpose, we implemented only such two functional blocks, as shown in Figure 8(b). The latency of these two blocks is 5 cycles, i.e. 20.8 ns, of the clock at 240 MHz. Specifically, the latency of the 20-to-30 converter is simulated to be 12.5 ns and that of the 30-to-20 converter is simulated to be 8.3 ns.

In order to measure the latency of each functional block, we also need to know the latency of the serializer and that of deserializer. These two parameters are not provided in the FPGA datasheet. These two latencies were obtained by measuring the time difference between the special input parallel pattern feeding into the serializer and the recovered special pattern on the receiver side, similar to what we did in Section 3. The latency of the serializer was measured to be 20.3 ns, and the latency of the deserializer was measured to be from 33.1 ns to 41.4 ns. By subtracting the latency of the known functional blocks described above from the full latency measurement, we obtained the latency of the encoder and the decoder to be 23.0 ns, which matches the simulated latency sum of the encoder (6.3 ns), the data extractor (12.5 ns), and the descrambler (4.2 ns), listed in Table 1. Note that the latency of the CRC examiner is not included.

As can be seen in the table, the latency of LOCx2 is less than that of LOCx2-130. The reason is that LOCx2 and LOCx2-130 operate at 320 MHz and 160 MHz, respectively. A time-critical path in the encoder design is the CRC calculation. In LOCx2, the pipeline technique is used and it takes two clock cycles, one of which is shared with the scrambling process, to calculate the CRC code. Due to the slower frequency, the scrambler, the CRC generator, and the frame builder can share a single clock cycle in LOCx2-130.

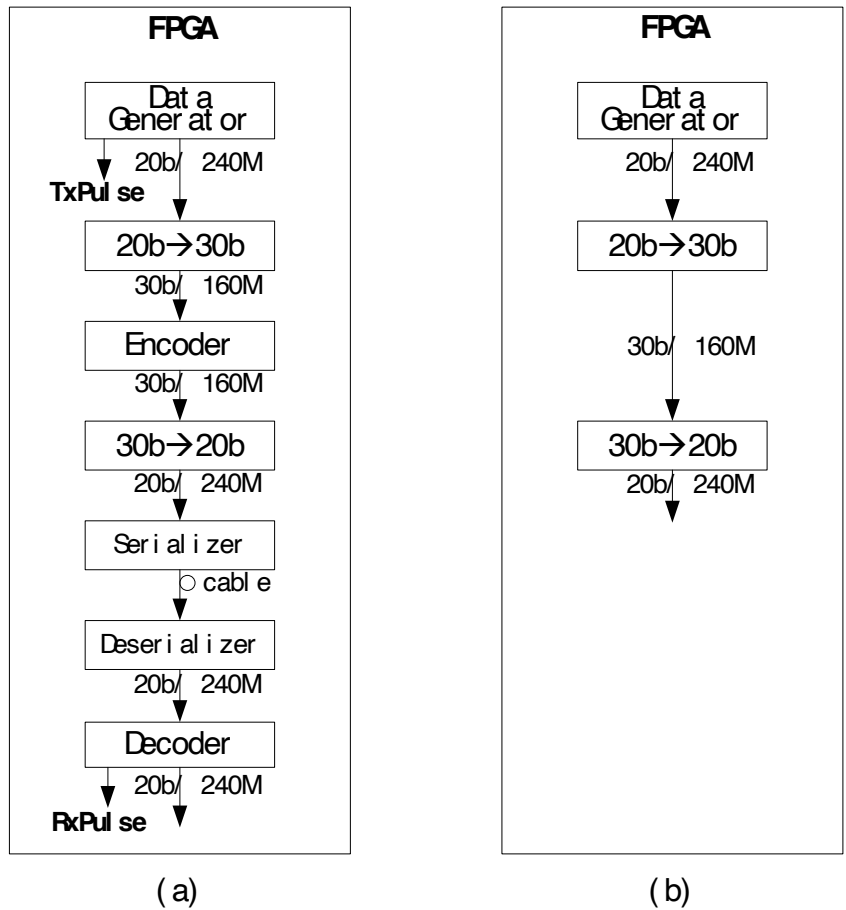

Figure 8: Block diagrams of the latency validation for the whole data link with a serializer and a deserializer (a) and 20-bit to/from 30-bit converters (b).

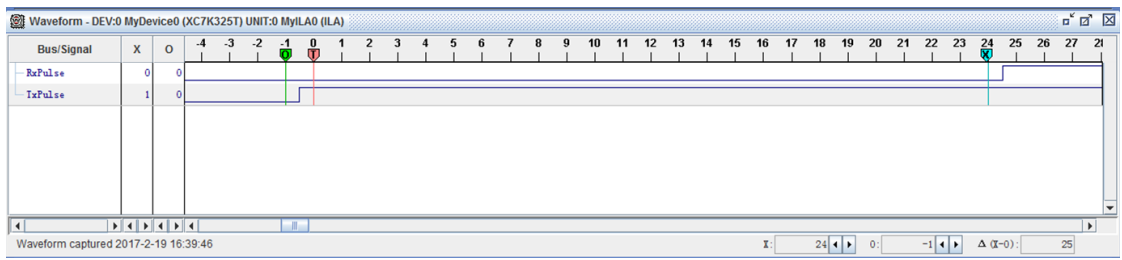

Figure 9. Latency simulation of the link shown in Figure 8(a). The latency from TxPulse to RxPulse is 25 clock cycles of 240 MHz, which is equal to 104.2 ns.

The reason that the latency of the whole link is not fixed comes from the FIFO and the deserializer. The latency of the FIFO depends on the phase difference between the ASIC internal clock and the ADC clock. The latency of the deserializer varies due the phase uncertainty of the parallel data clock, which is obtained by dividing the recovered clock of 5.12

GHz by 16 for LOCx2 or 4.80 GHz by 20 for LOCx2-130. The latency variation from the FIFO and the deserializer can be absorbed when multiple links are aligned into a single clock domain on the receiver side. The multiple-link alignment is beyond of the scope of this paper.

Table 1 Latency of each functional block in LOCx2 and LOCx2-130 data link (ns)

| Functional block | | | TX: LOCx2 RX: FPGA | | LOCx2 TX & RX implemented in FPGA | | TX: LOCx2-130 RX: FPGA | | LOCx2-130 TX & RX implemented in FPGA | |
|---|---|---|---|---|---|---|---|---|---|---|
| | | | Min | Max | Min | Max | Min | Max | Min | Max |
| 20B to 30B | | | 0.0 | 0.0 | 0.0 | 0.0 | 0.0 | 0.0 | 12.5 | 12.5 |
| TX | Encoder | FIFO | 8.4 | 11.6 | 3.1 | 6.3 | 16.4 | 22.7 | 0.0 | 0.0 |
| | | Scrambler CRC generator | 6.3 | 6.3 | 3.1 | 3.1 | 6.3 | 6.3 | 6.3 | 6.3 |
| | | Frame builder | 3.1 | 3.1 | 3.1 | 3.1 | | | | |
| | 30B to 20B | | 0.0 | 0.0 | 0.0 | 0.0 | 0.0 | 0.0 | 8.3 | 8.3 |
| | Serializer | | 6.3 | 6.3 | 14.4 | 14.4 | 11.7 | 11.7 | 20.3 | 20.3 |
| | Total | | 24.0 | 27.2 | 23.7 | 26.9 | 34.4 | 40.7 | 47.4 | 47.4 |
| RX | Deserializer | | 28.5 | 31.4 | 28.5 | 31.4 | 36.7 | 40.1 | 36.7 | 40.7 |
| | Decoder | Data extractor | 9.4 | 9.4 | 9.4 | 9.4 | 12.5 | 12.5 | 12.5 | 12.5 |
| | | Descrambler | 3.1 | 3.1 | 3.1 | 3.1 | 4.2 | 4.2 | 4.2 | 4.2 |
| | | CRC examiner | 3.1 | 3.1 | 3.1 | 3.1 | 4.2 | 4.2 | 0.0 | 0.0 |
| | Total | | 44.1 | 47.0 | 44.1 | 47.0 | 57.6 | 61.0 | 53.4 | 57.4 |
| Whole link | | | 68.2 | 74.3 | 67.8 | 73.9 | 92.0 | 101.7 | 100.8 | 104.8 |

## 5. Conclusion

We present the latency measurements of LOCx2 and LOCx2-130, including the latencies of the ASICs and their corresponding receivers implemented in the FPGA. The measurement results indicate that both ASICs achieve their design goals and meet the latency specifications. The measurements of ASIC latency match simulation. The latency of the whole link is consistent with that measured by using the Xilinx ChipScope Pro Analyzer tool. The consistency between the simulation and measurements validates the ASIC latency characterization.

## Acknowledgments

We acknowledge the support by the NSF and the DOE Office of Science, SMU's Dedman Dean's Research Council Grant, the National Natural Science Foundation of China under Grant No. 11705065 and Hubei Polytechnic University Key Research Project (16xjz04A). The authors are grateful to Drs. Jinhong Wang and Junjie Zhu from University of Michigan, Drs. Paulo Moreira, Syzmon Kulis, and Sandro Bonacini from CERN, Drs. Hucheng Chen, Kai Chen, and Hao Xu of Brookhaven National Laboratory.